# AMBER: un instrument focal proche infrarouge pour le VLTI

# AMBER: a near infrared focal instrument for the VLTI

Romain G. Petrov[*a], Fabien Malbet[b], Andrea Richichi[c], Karl-Heinz Hofmann[d], Denis Mourard[e], Karim Agabi[a], Pierre Antonelli[e], Eric Aristidi[a], Carlo Baffa[c], Udo Beckmann[d], Philippe Berio[e], Yves Bresson[e], Frederic Cassaing[f], Alain Chelli[b], Albrecht Dreiss[d], Michel Dugué[e], Gilles Duvert[b], Thierry Forveille[b], Sandro Gennari[c], Michael Geng[d], André Glentzlin[e], Daniel Kamm[e], Stéphane Lagarde[e], Etienne LeCoarer[b], Danielle LeContel[e], Franco Lisi[c], Bruno Lopez[e], Alessandro Marconi[c], Gilbert Mars[e], Grégoire Martinot-Lagarde[g], Djamel Mekarnia[e], Jean-Louis Monin[b], David Mouillet[b], Pascal Puget[b], Yves Rabbia[e], Sylvestre Rebattu[e], François Reynaud[h], Sylvie Robbe-Dubois[a], Karine Rousselet-Perraut[b], Michel Sacchettini[b], Piero Salinari[c], Markus Schoeller[j], Isabelle Tallon-Bosc[i], Martin Vannier[a], Gerd Weigelt[d].

[a] Université de Nice–Sophia Antipolis, Parc Valrose, F-06108 Nice Cedex 02, France;
[b] Observatoire de Grenoble, Université Joseph Fourier, BP53, F-38041 Grenoble Cedex, France;
[c] Osservatorio Astrofisico di Arcetri, Largo E. Fermi 5, I-50125 Firenze, Italia;
[d] Max Planck Institute für Radiosatronomie, Auf den Hügel 69, D-53121 Bonn, Deutschland;
[e] Département Fresnel, Observatoire de la Côte d'Azur, BP 4229, F-06304 Nice Cedex 04, France;
[f] Office National d'Etudes et de Recherche Aérospatiales, BP 72, F-92322 Châtillon Cedex, France;
[g] Institut National des Sciences de l'Univers, BP 287, F-75766 Paris Cedex 16, France;
[h] Institut de Recherche en Communications Optiques et Microondes, F-87060 Limoges Cedex, France;
[i] Centre de Recherche Astronomique de Lyon, F-69561 Saint Genis-Laval Cedex, France;
[j] European Southern Observatory, Karl-Schwarzschild-Sts.2, D-85748 Garching, Deutschland.

**RESUME**

AMBER est l'instruments focal proche infra rouge du mode interférométrique du Très Grand Télescope Européen. Les spécifications de cet instrument généraliste ont été définies pour trois programmes clefs sur les systèmes stellaires en formation, les régions centrales des noyaux actifs de galaxies et les masses et les spectres de planètes extra solaires géantes chaudes. Il combine trois faisceaux, ce qui lui donne une capacité de reconstruction d'images, et est optimisé pour la précision des mesures grâce à un filtrage spatial des faisceaux par des fibres optiques mono mode et à une combinaison des informations obtenues simultanément à différentes longueurs d'onde.

**ABSTRACT**

AMBER is the General User near infrared focal instrument of the Very Large Telescope Interferometer. Its specifications are based on three key programs on Young Stellar Objects, Active Galactic Nuclei central regions, masses and spectra of hot Extra Solar Planets. It has an imaging capacity because it combines up to three beams and very high accuracy measurement are expected from the spatial filtering of beams by single mode fibers and the comparison of measurements made simultaneously in different spectral channels.

---

[*] Correspondence: e_mail: petrov@unice.fr; Telephone: 33 4 92 07 63 47; Fax: 33 4 92 07 63 21.

# 1. INTRODUCTION

## 1.1 THE VERY LARGE TELESCOPE INTERFEROMETER

The Very Large Telescope Interferometer (VLTI) [1] will eventually combine the beams produced by any subset of the seven (or eventually more) telescopes present on the Paranal mountain in Chile. Each beam will be partially corrected from atmospheric wave front perturbation thanks to Adaptive Optics (AO) modules for the four 8 m Unit Telescopes (UT) or to tip-tilt correctors for the three 1.8 m Auxiliary Telescopes (AT). The beams will be transported to the interferometric laboratory through the telescope Coudé trains feeding Delay Lines installed in a thermally stable interferometric tunnel. In a first step, the OPD will be set to zero with errors smaller than 100 µm thanks to a good global metrology of the instrument. Later, the inclusion of fringe sensors[2] in the OPD control loop will allow stabilizing the fringes within a small fraction of wavelength. For each beam the pupils and the images delivered in the focal laboratory are stabilized. Eventually, a field separator implemented at each telescope will allow AO correction and fringe tracking on a reference star up to 1 arc minute away from the science source. The ultimate feature of the VLTI will be a metrology system combined with differential delay lines allowing to perform imaging through phase referencing between the off axis reference star and the science source as well as high accuracy differential astrometry. The implementation of the VLTI will be progressive. It will start with only two telescopes without fringe tracking and eventually reach the full picture evoked above. The VLTI has to be completed by focal instruments which can clean or calibrate perturbations in each beam and then combine two or more beams and record and analyze the interference fringes in one or several spectral channels.

## 1.2 STRATEGIC CHOICES BEHIND THE AMBER CONCEPT AND SPECIFICATIONS

AMBER is a near infrared, three beams, dispersed fringes, single mode VLTI focal instrument.

Besides a very wide scientific program, the near infrared, and particularly the K band, offers the possibility to work with nearly diffraction limited images, thanks to tip-tilt correction on the ATs and a modest order Adaptive Optics system on the UTs. This allows to apply techniques well proven on smaller aperture interferometer without the necessity to correct for the rapid temporal variation of the thermal background like in the mid infrared instrument MIDI[3].

AMBER must combine three beams to be the first VLTI instrument with some imaging capability.

However, since image reconstruction with three telescopes will be very slow, most of the astrophysical information is to be expected from the confrontation of models with a limited number of measures. In such a model fitting, analyzing the measures as a function of wavelength increases very significantly the number and the quality of measures and then the constraints imposed to the models. This pointed to a dispersed fringes instrument, similar to what has been experienced on the Plateau de Calern interferometers in France [4].

The very strong demand for accuracy implied two other strategic choices. The first one was to give priority to "single mode operation" where each beam is filtered in the image plane by a spatial filter (very small pinhole or single mode fiber) which transmits only one coherent mode. After the filter, the beam has a single phase and all atmospheric noise is reduced to OPD noise, which can be frozen, and to photometric fluctuations which can be monitored. This is similar to what has been used in the Fluor/IOTA[5] or in the PTI[6] interferometers. The second one was an even stronger motivation to implement dispersed fringes to perform "differential interferometry" which measures the differences in visibility and/or phase between two spectral channels. GI2T[4] and in Differential Speckle Interferometry [7] has shown that differential measurement are at least 10 times more accurate than absolute measurements, almost whatever the quality of the latest.

## 1.3 THE AMBER CONSORTIUM

The Institutes and Laboratories in charge of the study, construction, commissioning and hand over to ESO of AMBER are the first five in the authors affiliation list of this paper. The warm optics, from the interface with the VLTI to the interface with the cooled spectrograph, the static and remotely controlled mechanics, the electronics hardware and the instrument control software are built in Nice. The cooled spectrograph with its optics, static and remotely controlled mechanics is produced in Florence. The Detector, its electronics, the data acquisition chain and the real time processing are provided by Bonn. The software to prepare, manage and reduce the observations is carried out in Grenoble, where the global integration and laboratory tests in Europe will be performed. Individual participants from the other Institutes in the affiliation list provide expertise in optical interferometry as well as on the astrophysical objectives of AMBER.

## 2. GENERAL SPECIFICATIONS AND CHARACTERISTICS OF AMBER

### 2.1 SCIENCE DRIVERS FOR THE AMBER REQUIREMENTS

AMBER is a "General User" instrument. Therefore, its scientific program includes all possible programs with an interferometer working in the near infrared with resolutions reaching one milliarcsecond (mas)[8]. A non exhaustive list is summarized in the following table which gives estimates of the requirements needed for breakthrough research in each field.

| Topic | Maximum error on the visibility and/or the differential phase (rad) | Minimum K magnitude | Spectral Coverage | Spectral Resolution |
|---|---|---|---|---|
| Extra solar planets | $10^{-4}$ | 5 | J+H+K | 35 |
| AGN dust tori | $10^{-2}$ | 11 | K | 35 |
| QSO and AGN BLR | $10^{-2}$ | 11 | J,H,K | 1000 |
| Young Stellar Objects | $10^{-2}$ | 7 | J,H,K, lines | 1000 |
| Circumstellar material | $10^{-2}$ | 4 | J,H,K, lines | 1000 |
| Binaries | $10^{-3}$ | 4 | K | 35 |
| Stellar Structure | $10^{-4}$ | 1 | lines | 10000 |

Table 1: Scientific programs and requirements used to establish and check the specifications of AMBER

• Young Stellar Objects (multiplicity, disks and jets scale and structure) and protoplanetary disks [9],[10],[11]. Hundreds of candidates are accessible if AMBER achieves a visibility accuracy better than 1% on K>7 stars. Among the programs of high astrophysical interest, this is the "easy" one and is therefore considered as the minimum target.

• AGN: Dust tori and jets scale and structure can be constrained by visibility and phase closure measurements in the near-IR continuum [12]. Differential interferometry in the near-IR emission lines should provide angular and kinematic information on the Narrow Line and Broad Line Regions thanks to the capacity of this technique to resolve structures much smaller than the diffraction limit. A feasibility study made within the AMBER Science Group[13] shows that for the bright QSO 3C273.0 we could constrain the size of its BLR and locate the photocenter of 300 km/s radial velocity bins with a 10 mpc accuracy (i.e. about 1% of the BLR radius) corresponding to a $5\ 10^{-3}$ radians uncertainty on the differential phase. This requires only the capacity to track the fringes and sense the wave fronts on objects with magnitudes K between 10 (2 or 3 targets) and 13 (more than 70 targets).

• Extra solar planets: direct detection, orbits, masses and low resolution spectroscopy (yielding the temperature and constraining the atmosphere composition) of hot massive planets such as 51 Peg or any Jupiter mass planet less than 0.1 or 0.2 au away from a sun like stars a 10 pc [14],[15].

• Circumstellar material: morphology and kinematics of the material around hot and cold[16] stars

- Binaries [11]: main sequence and giant stars, brown dwarfs [15], interaction with the circumstellar material, etc…

- Stellar structure and activity: limb-darkening, rotation, stellar spots, non radial pulsation, etc…

## 2.2 BASIC CONCEPT OF AMBER

Figure 1 summarizes the key elements of the AMBER concept.

AMBER has a multi axial beam combiner. A set of collimated and parallel beams are focussed by a common optical element in a common Airy pattern which contains fringes (-1- in figure 1). The output baselines are in a non redundant set up, i.e. the spacing between the beams is selected for the Fourier transform of the fringe pattern to show separated fringe peaks at all wavelengths. The Airy disk needs to be sampled by a lot of pixels in the baseline direction (an average of 4 pixels in the narrowest fringe, i.e. at least 12 pixels in the baseline direction) while in the other direction only one pixel is sufficient. To minimize detector noise each spectral channel is concentrated in a single column of pixels (-3- in figure 1) by cylindrical optics (-2- in figure 1). This multi axial beam combiner has been selected because it allows an easy evolution from two to three or even 4 telescopes. A second advantage is that it makes it easy to fit a spectrograph. It has been shown that for a given sampling of the fringes, the number of pixels used in this multi axial scheme is comparable to the product of the number of measures by the number of pixels used in a co-axial beam combiner.

The fringes are dispersed by a standard "long slit" spectrograph (-4- in figure 1) on a two dimensional detector (-5- in figure 1). For work in the K band with resolutions up to 10 000 the spectrograph must be cooled down to about –60°C with a cold slit in the image plane and a cold pupil stop. In practice we found it simpler to cool it down to liquid nitrogen temperature. This combination of a multi axial beam combiner with a spectrograph has been used for years in the French interferometers on the Plateau de Calern [4].

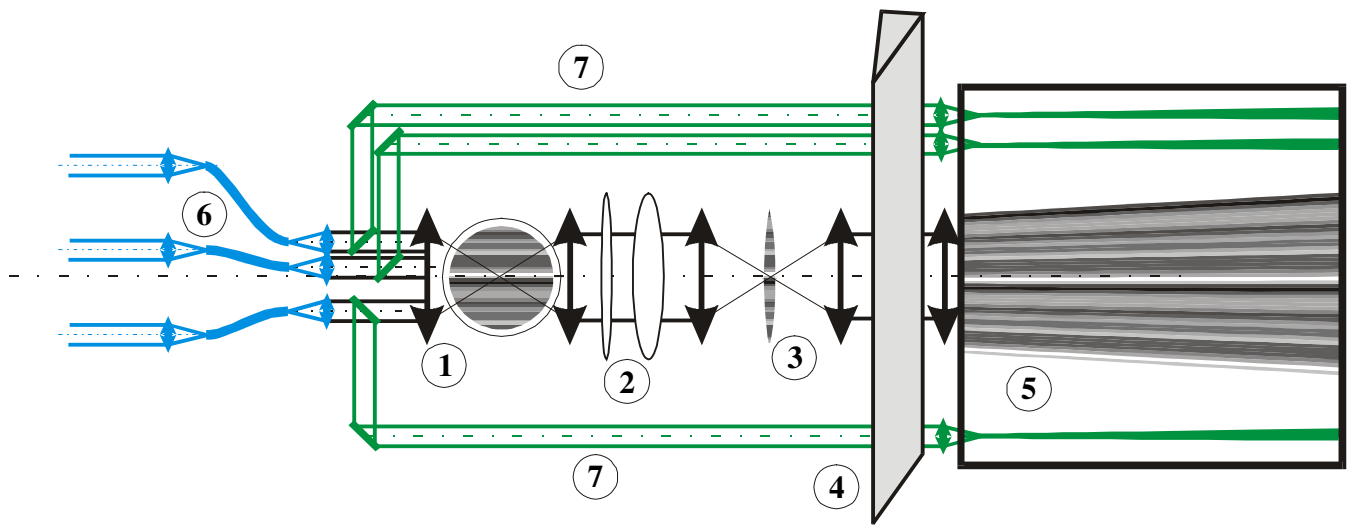

Figure 1: Basic concept of AMBER: (1) multi axial beam combiner, (2) cylindrical optics, (3) anamorphosed focal image with fringes, (4) "long slit spectrograph", (5) dispersed fringes on 2D detector, (6) spatial filter with single mode optical fibers, (7) photometric beams.

To produce high accuracy measurements, it is necessary to spatially filter the incoming beams to force each one of them to contain only a single coherent mode. To be efficient, the spatial filter must transmit at least $10^3$ more light in the guided mode than in all the secondary modes. For the kind of imperfect AO correction (Strehl ratios often < 50%) available for the VLTI, the single way to achieve such high filtering quality with decent light transmission is to use single mode optical fibers [17] (-6- in figure 1). The flux transmitted by each filter must be monitored in real time in each spectral channel. This explains why a fraction of each beam is extracted before the beam combiner and send directly to the detector through a dispersive element (-7- in figure 1). The instrument must also perform some beam "cleaning" before entering the spatial filter, such as correcting for the differential atmospheric refraction in the H and J bands or, in some cases, eliminating one polarization.

## 2.3 AMBER MEASUREMENTS

For each elementary frame and in each AMBER spectral channel, i.e. in each column of the detector, we have [18] an interferometric signal and three photometric signals which have been processed by the same optics and the same dispersive elements.

A generalization of the ABCD algorithm used in the PTI interferometer [6] has been developed to establish a linear relation between the values measured in each pixel and the complex coherence for each baseline. The "pixel to visibility matrix" (PTVM) is calibrated using the artificial source unit in the instrument. This matrix combines the effects of many parameters which affect the measurement such as the detector gain table and the exact shape of the beams after the fiber output and the pupil stop in the spectrograph. To calibrate this matrix we must combine measures of the full three beams interferogram, measures of the signals detected when only one beam is open and measures of the interferograms obtained when one of the beams is affected by a known delay close to $\pi/4$. To achieve the goal on a visibility uncertainty smaller than $10^{-4}$ this calibration phase delay must be reproducible with a $10^{-4}$ radians accuracy.

For each baseline l-m, we obtain in each spectral channel, a measure of the visibility $V_{lm}(\lambda)$ and of the phase $\Phi_{lm}(\lambda)$. This is used to derive the various parameters measured by AMBER.

### 2.3.1 Calibrated visibility as a function of wavelength $V_{lm}(\lambda)$.

After time averaging and various debiasing and calibration steps [19] (correction of residual piston effects using the fringe tracker residuals, use of a star with known visibility…), $V_{lm}(\lambda)$ yields the calibrated visibility in each spectral channel for the baseline l-m.

### 2.3.2 Differential visibility as a function of wavelength $V_{lm}(\lambda)/V_{lm}(\lambda_0)$

One or several spectral channels can be defined as being a reference channel (when several wavelengths are used, the information is averaged over $\lambda$). Then the ratio between the visibility in each spectral channel and the visibility in the reference channel yields the differential visibility $V_{lm}(\lambda)/V_{lm}(\lambda_0)$. The interest of this quantity is that it can be calibrated more accurately than the absolute visibility alone because many effects will affect the visibility in many channels in the same way (or in ways related by known relations).

### 2.3.3 Differential phase as a function of wavelength $\Phi_{lm}(\lambda) - \Phi_{lm}(\lambda_0)$

The absolute phase $\Phi(\lambda)$ has no meaning unless the phase referencing [1] equipment is available. However, it is immediately possible to measure the absolute phase differences between any spectral channel and the reference channel: $\Phi_{lm}(\lambda) - \Phi_{lm}(\lambda_0)$. The measured phase difference $\Delta\Phi_{lm}(\lambda)$ will be the sum of:

$$\Delta\Phi_{lm}(\lambda) = \Delta\Phi_{lm*}(\lambda) + \Delta\Phi_{lma}(\lambda) + \Delta\Phi_{lmI}(\lambda)$$

where $\Delta\Phi_{lm*}(\lambda)$ is due to the source, $\Delta\Phi_{lma}(\lambda)$ is the contribution of the differential atmospheric refraction and $\Delta\Phi_{lmI}(\lambda)$ is the instrumental chromatic phase difference between the beams l and m. The dominant and most difficult to eliminate term is the instrumental one $\Delta\Phi_{lmI}(\lambda)$. AMBER plans to solve this problem by inverting the beams at the entrance of the interferometric laboratory, or in any case before any chromatic optics in AMBER (dichroics, fibers, beam splitters, cryostat windows, spectrograph chamber, detector…). Then we obtain a new phase difference $\Delta\Phi_{ml}(\lambda)$:

$$\Delta\Phi_{ml}(\lambda) = -\Delta\Phi_{lm*}(\lambda) - \Delta\Phi_{lma}(\lambda) + \Delta\Phi_{lmI}(\lambda).$$

The difference eliminates the instrumental term $\Delta\Phi_{lmI}(\lambda)$. The atmospheric term $\Delta\Phi_{lma}(\lambda)$ can be measured on a reference star given that the small difference in zenith angle and in pressure and temperature inside the tunnel are corrected for. These two effects result in a single multiplication factor which can be fitted in the data itself (or ideally estimated from spectral channels where the astrophysical contribution $\Delta\Phi_{lm*}(\lambda)$ is known to be negligible). This beam commutation operation is the key to measure extremely accurate differential phases such as the one needed for the study of extra solar planets.

**2.3.4 Phase closure $\Phi_{123}(\lambda)$.**

In the phase closure relation $\Phi_{123}(\lambda) = \Phi_{12}(\lambda) + \Phi_{23}(\lambda) + \Phi_{31}(\lambda)$ all atmospheric and most of instrumental terms are cancelled. The only terms which are not corrected in the phase closure result directly from the fringe detection process (error in the PTV matrix).

For the extra solar planet application, it might be interesting to use the differential phase closure [15]. To obtain a measure independent from any instrumental influence, we are investigating a solution based on small spectral shifts between two exposures. We then measure successively:

$$\Phi_{123}(\lambda) = \Phi_{123*}(\lambda) + \Phi_{123M}(\lambda)$$

$$\Phi_{123}(\lambda+\delta\lambda) = \Phi_{123*}(\lambda+\delta\lambda) + \Phi_{123M}(\lambda+\delta\lambda)$$

where $\Phi_{123*}(\lambda)$ is due to the star and $\Phi_{123M}(\lambda)$ is an error term due to the measurement process (gain table variation, etc…). If we assume that, when all the spectral calibrations are made and when the wavelength shift $\delta\lambda$ is small we have $\Phi_{123M}(\lambda) = \Phi_{123M}(\lambda+\delta\lambda)$, then we measure the derivative of the phase closure as a function of $\lambda$ without any instrumental bias:

$$[\Phi_{123}(\lambda) - \Phi_{123}(\lambda+\delta\lambda)]/\delta\lambda = [\Phi_{123*}(\lambda) - \Phi_{123*}(\lambda+\delta\lambda)]/\delta\lambda$$

**2.4 OBSERVING AND CALIBRATION CYCLES OF AMBER**

AMBER has only one observing mode. In each case we read frames (the individual read out of the detector yielding pixel values in all spectral channels) which are combined in exposures (successive frames recorded on the same source with the same instrument set up). Each exposure yields measurements of the complex coherence but a complete calibration needs combining exposures on different sources (science target, reference star, sky background, calibration lamps…) or with different set ups (repeating the same exposure after a beam commutation or a small shift in wavelength) [19].

The frame time is selected according to the science requirements and the availability of the fringe tracker. For very accurate measurements on relatively bright objects, the frame time will be set to a minimum value of 10 ms (necessary to read about all the spectral channel in the K band at the lowest resolution). This is necessary to freeze almost completely all atmospheric and instrumental OPD fluctuations, even if the fringes are stabilized by a fringe tracker. For observing the fainter objects, the frame time can be extended up to 100 ms. The accuracy will be degraded by the seeing dependant contrast loss due to imperfectly frozen OPD fluctuations. In this mode, the absolute visibility accuracy will hardly be better than 1%. It is worth remembering that the differential visibility and phase should not be biased by this extended frame time. When the fringes are stabilized by a fringe tracker, the frame time can be extended from 1 to 100 s. The 100 s limit is set by the variation of the projected baseline with time. In practice, we will use the shortest frame time allowing the photon noise to clearly dominate the detector read out noise. This longer frame times are necessary for most high spectral resolution observations.

The exposure time (or the number of frames in one exposure) is set for a given object by the desired accuracy as a function of the magnitude, the instrument set-up, the average seeing conditions.

The exposure cycle results from the AMBER measurement which is considered as having the highest priority for a given science objective. For visibility measurements, the crucial step is to use a reference star with known visibility. Then the exposure cycle will contain a sequence of exposure on calibration lamps (the PTV matrix calibration), sky, science and reference source, with a priority to reduce the time separating the associated exposures on science and reference sources down to 5 minutes. For differential phase measurements, the exposure cycle will contain in addition successive exposures with an without a beam commutation. The priority is to have this exposures separated by less than 30 seconds. Reference sources will continue to be used but with a smaller frequency.

All operations needed to perform a given measurement, i.e. a given exposure cycle are automatically unrolled by the instrument. The sequences are described in standard templates, with parameters set by the User as a function of its scientific objectives. To each given exposure cycle corresponds a standard way to reduce the data to calibrated measurements. However, all data will be recorded to allow more sophisticated post processing.

**2.5 SUMMARY OF AMBER TOP LEVEL SPECIFICATIONS**

Table 2 summarizes the top level specifications of AMBER after the various tradeoffs made during the design phases.

| Characteristic | Specification | | | Goal (or expected performance) | | |
|---|---|---|---|---|---|---|
| Number of beams | 3 | | | 3 | | |
| Minimum spectral resolution | 35 in K | | | ≈ 35 in J | | |
| Medium spectral resolution in K | $500 < \Re < 1000$ | | | | | |
| Highest spectral resolution in K | | | | $10\,000 < \Re < 15\,000$ | | |
| Spectral resolution in H and J | As it results from the K band equipment. Use order 2 in J. | | | | | |
| Spectral coverage | J,H,K' from 1 to 2.3 µm | | | J,H,K from 1 to 2.4 µm | | |
| Instantaneous spectral coverage | Simultaneous observation of the full spectral domain for $\Re=35$ | | | | | |
| Absolute visibility accuracy | $3\sigma_V=0.01$ | | | $\sigma_V=10^{-4}$ | | |
| Differential visibility accuracy | $3\sigma_{(V/V0)}=10^{-3}$ | | | $3\sigma_{(V/V0)}=10^{-4}$ | | |
| Differential phase accuracy | $3\sigma_{\Delta\Phi}=10^{-3}$ rad | | | $\sigma_{\Delta\Phi}=10^{-5}$ rad | | |
| Limiting magnitude for 5σ fringe detection (V=1) | K=11 | H=11 | | K=13 | H=12.5 | J=11.5 |
| Limiting magnitude with off-set reference star | | | | K=19.5 | H=20 | J=18.5 |
| Instrument contrast | 0.8 | | | 0.9 | | |
| Optical throughput (optics, fibers, spectro, detector) | 2% in K | 1% in H | 1% in J | 5% in K | 5% in H | 5% in J |
| Instrument contrast stability | $10^{-2}$ over 5 minutes | | | $10^{-3}$ over 5 minutes | | |
| Differential phase stability | $10^{-3}$ rad over 1 minute | | | $10^{-4}$ rad over 1 minute | | |

Table 2: summary of the top level specifications of AMBER

The goals given for the limiting magnitudes, in lines 11 and 12 of table 2, are actually our estimate of what will be the performances of AMBER given the optical throughput and instrument contrast estimated from the actual implementation of AMBER at the end of the design phase [20]. The computation [21] assumed the use of a single polarization and a Strehl ratio in K for a on axis reference source = 0.5 (when the science and reference sources are 1 arc minute away, the Strehl in K is divided by two). Without fringe tracking, the frame time was taken equal to 100 ms (high sensitivity mode). With off axis fringe tracking, the limiting magnitude was computed for a 10% visibility accuracy obtained after a 4 hour exposure divided in 144 frames of 100 s.

## 3. OVERVIEW OF THE AMBER IMPLEMENTATION

Figure 2 shows the global implementation [20] of AMBER with the additional features needed by the actual operation of the instrument.

There are three spatial filters, one for each spectral band, because of the limited wavelength range over which a fiber can remain single mode. The three spatial filters inputs are separated by dichroic plates. After the fiber outputs, a symmetric cascade of dichroics combines the different bands again, but the output pupil in each band has a shape proportional to the central wavelength of the band. Therefore the Airy disk and the fringes have the same size for all central wavelength. This allows to use the same spectrograph achromatic optics for all bands and to have the same sampling of all the central wavelengths.

Then the beams enter the cylindrical optics anamorphoser "OPM-ANS" before entering the spectrograph through a periscope used to align the beam produced by the warm optics and the spectrograph. The spectrograph has an image plane cold stop (SPG-INW), a wheel with pupil masks for 2 or 3 telescopes and spectral filters (SPG-FMW). The separation between the interferometric and photometric beams is performed in a pupil plane inside the spectrograph (SPG-IPS), after the image plane cold stop. After dispersion, the spectrograph chamber (SPG-CHA) sends the dispersed image on the detector chip.

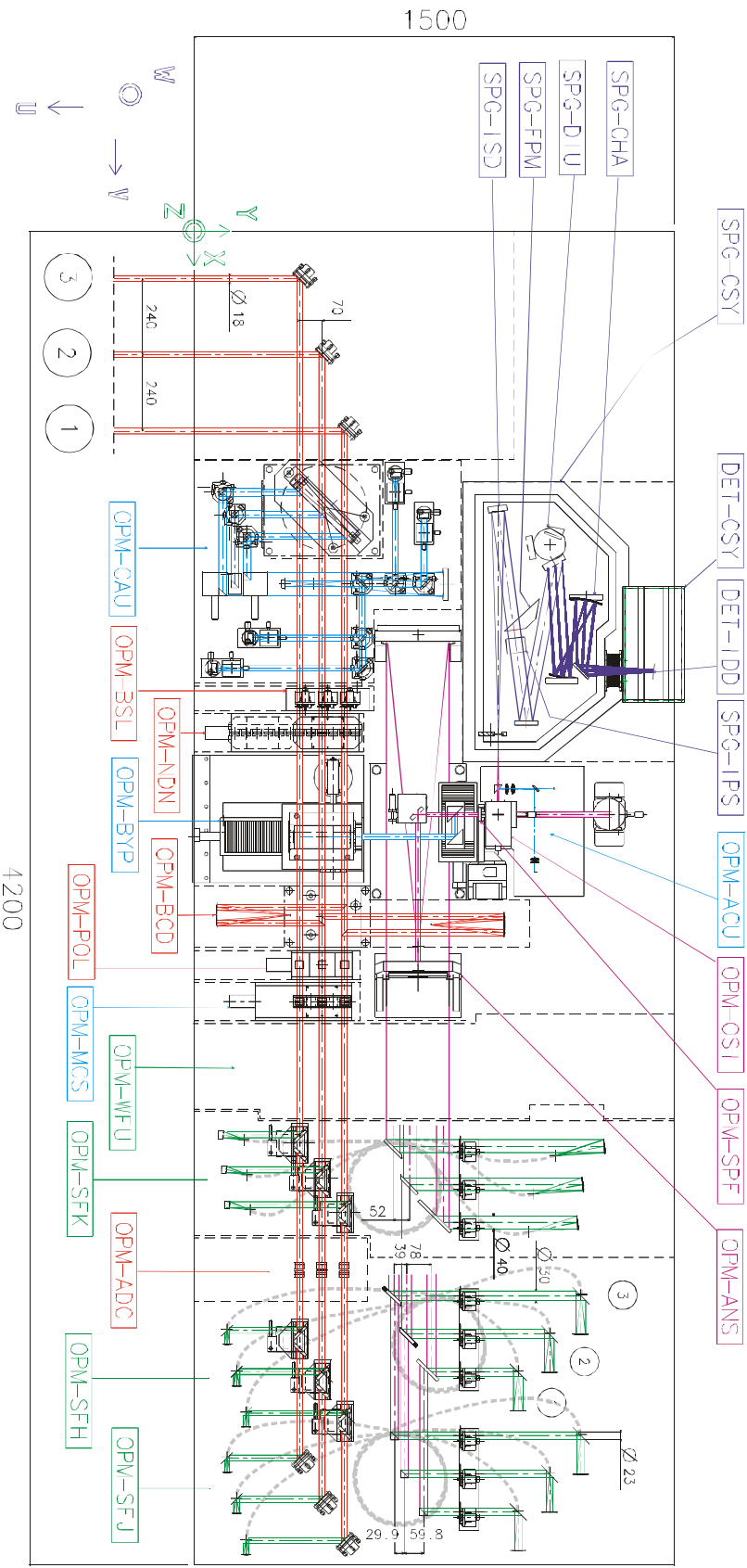

Figure 2: General Implementation of AMBER. The VLTI beams arrive in the lower left corner. OPM-CAU: calibration and alignment unit. OPM-BCD: beam inverting device. OPM-POL: movable polarisation selecting device. OPM-SFK, OPM-SFH, OPM-SFJ: spatial filters for the K, H and J bands. OPM-ADC: corrector for the atmospheric diffential refraction in the H and J bands. OPM-ANS: cylindrical afocal system for image ananmorphosis. OPM-OSI: periscope to co alignm the warm and the cold optics. OPM-ACU: Alignment control unit: a visible CCD and a single pixel IR detector control pupil and image positions, alignment and optical quality. OPM-BYP: movable bypass feeding directly the VLTI beams in the spectrograph or in the ACU to check VLTI alignment and acquire complex fields. The cooled spectrograph has an image plane cold stop:SPG-INW, a wheel of pupil plane cold stops and filters: SPG-FMW, allows the separation between interferometric and photometric beams: SPG-IPS, disperses the light: SPG-DIU and its chamber SPG-CHA sends the dispersed fringes on the Hawaii detector in a separated cryostat: DET-CSY. During final operation, the two cryostats are connected by a cold tunnel and share the same vacuum...................................................

The Calibration and Alignment Unit (OPM-CAU), contains all calibration lamps and can emulate the VLTI in the integration and test phase. A set of plane parallel plates can be introduced in the beam in order to introduce the $\lambda/4$ delays in one beam necessary to calibrate the matrix of the "pixel to visibility" linear relation. Just before the entrance of the spectrograph, the beam can be fed in a Alignment Control Unit (OPM-ACU) which has a visible technical CCD and a single pixel IR detector to check that the instrument is correctly aligned and if the fibers are correctly fed. This unit can also control the positions of the pupils and analyze a certain field around the center of the spatial filter thanks to a movable "bypass" allowing to feed the beams coming from the VLTI directly inside the OPM-ACU detectors or the spectrograph.

Several components of the AMBER instrument, such as the dichroics, the fibers, the filters, the beam splitter, the cryostat window are optimized for only one polarization. Then, the other polarization will provide only a small gain in flux but can produce a substantial loss in contrast. To avoid this, one polarization can be eliminated by movable polarization cubes (OPM-POL).

The OPM-BCD is a tentative representation of the Bean Commuting Device. It is necessary to commute the beams without inverting the individual images, to avoid readjusting the atmospheric refraction corrector (OPM-ADC) at each beam inversion. We still hope that the beam commuter will be implemented by ESO in the focal laboratory before the separation between AMBER, MIDI and the Fringe Tracker. This would allow measuring accurately the phase differences between these instruments and then increasing the wavelength range for differential measurements.

At a later phase, a wide field unit (OPM-WFU) will bypass the spatial filters in a way which does not destroy the interferometric beam combination (the output pupils will have the same centers than the spatial filters ones) to allow multiple mode operation, which is probably the best way to obtain interferometric information on objects which are resolved by the individual apertures but have features of size comparable to the interferometer resolution [22]. This non filtered beams can also be used to feed a future extension of AMBER to visible wavelength, around the $H_\alpha$ line, or to longer infrared wavelengths such as the L band. Such an extension would be installed on a "second floor" above the present table.

## 4. CONCLUSION

We are now close to the "Final Design Revue" of AMBER. All optical, mechanical and software studies performed so far confirm that AMBER will reach its basic specifications with a substantial margin. It is too early to say if we will be able to achieve our most ambitious goal which is the spectroscopy of hot extra solar giant planets, but we still have not found any show stopper.

In December 2001 we should ship to Paranal the instrument containing only the K band spatial filter (and all the "service" modules). Commissioning with the Siderostats shall take place between February and May 2002. Then we will commission the instrument for two telescopes as soon as two ATs are available (October 2002). The period between the end of the siderostat commissioning of the K band instrument and the availability of two AT will be used to install and commission the other spatial filters. The acceptance of AMBER with two ATs should take place in early 2003, which corresponds to the opening of AMBER to the community. Commissioning with two UTs will take place as soon as two AO systems have been delivered and commissioned for interferometry (first semester 2003). Three telescopes commissioning should begin in the first semester of 2003, as soon as 3 ATs are available. The full hand over to ESO of the three telescopes, three spectral bands instrument should take place in the second semester 2003.

Sometime in 2003 or 2004, it will be possible to use the dual field VLTI facility, first to correct wave fronts and track fringes on off axis reference stars, which will open the magnitude range between $K \approx 12$ and $K \approx 20$ for sources who are less than 1 arc minute away from a bright (K<13) reference star. In 2004, the accurate metrology between the VLTI science and reference beam will allow AMBER to use phase reference imaging, which is likely to speed up substantially the image reconstruction process.

However, using the full imaging potential of the VLTI, will require a second generation instrument, able to combine seven or height beams. AMBER as a limited capacity for an extension of its spectral coverage toward the visible or the L band. If we decide and manage to kick off the work for one of this extensions before the end of 2001, it could be implemented as soon as 2004 and use some of AMBER service modules and most of the instrument and observation control software.